# OPTIMAL INVESTMENT AND CONSUMPTION IN A BLACK–SCHOLES MARKET WITH LÉVY-DRIVEN STOCHASTIC COEFFICIENTS

BY ŁUKASZ DELONG[1] AND CLAUDIA KLÜPPELBERG

*Warsaw School of Economics and Munich University of Technology*

In this paper, we investigate an optimal investment and consumption problem for an investor who trades in a Black–Scholes financial market with stochastic coefficients driven by a non-Gaussian Ornstein–Uhlenbeck process. We assume that an agent makes investment and consumption decisions based on a power utility function. By applying the usual separation method in the variables, we are faced with the problem of solving a nonlinear (semilinear) first-order partial integro-differential equation. A candidate solution is derived via the Feynman–Kac representation. By using the properties of an operator defined in a suitable function space, we prove uniqueness and smoothness of the solution. Optimality is verified by applying a classical verification theorem.

**1. Introduction.** A fundamental problem in financial mathematics is the allocation of funds between assets in order to provide sufficiently large payments during the duration of an investment contract, as well as to arrive at a high return at maturity. This optimization problem has its origin in a seminal paper by Merton [18], where it is formulated as a utility maximization problem and an optimal strategy is derived via the Bellman equation. Since then, there has been a growing interest in investment and consumption problems and the classical Merton problem has been extended in many directions. One of the generalizations considers financial coefficients (risk-free return, drift and volatility) affected by an external stochastic factor.

Received February 2007; revised September 2007.
[1]Supported in part by an Advanced Mathematical Methods for Finance (AMaMeF) grant from European Science Foundation and by the Ministry of Education in Poland Grant M111 002 31/0165.

In this paper, we extend the results from [5] and [17]. We investigate a Black–Scholes-type financial model with coefficients depending on a background driving process. The dependence is described through general functions which satisfy linear growth conditions. An external stochastic factor is chosen as Ornstein–Uhlenbeck process driven by a subordinator. The Barndorff-Nielsen and Shephard model considered in [5] and [17] arises as a special case. As an additional possibility, the investor is allowed to withdraw (consume) funds during the term of the contract. This leads to an optimal investment and consumption problem which is more complex than a pure investment problem. From an analytical point of view, the difference is that, after applying the usual separation of variables, we arrive at a nonlinear partial integro-differential equation, whereas [5] and [17] deal with a linear one.

The first goal of this paper is to show that a candidate value function is the classical solution of a corresponding Hamilton–Jacobi–Bellman equation. This requires proving the existence of a classical solution to a nonlinear (semilinear) first-order partial integro-differential equation. It is well known (see [9], Chapter 12.2, and [20]) that the regularity of solutions to equations with an integral term is uncertain, especially in the degenerate case. There exist some results concerning the smoothness of a solution to a linear partial integro-differential equation (see, e.g., [2], Chapters 3.3 and 3.8, [9], Chapter 12.2 and [20]), but they all deal only with the nondegenerate second-order case. The degenerate case can be handled by applying a viscosity approach (e.g., [9], Chapter 12.2) which we want to avoid, following instead [5], where the existence of a classical solution to a linear first-order partial integro-differential equation is established. We believe that our proof (in Sections 4 and 5) of the existence of a unique classical solution to a nonlinear first-order partial integro-differential equation contributes to the present state of the literature.

Our second goal is to provide an explicit formula for the optimal consumption. In the case of a power utility function, it is intuitively easy to foresee a formula for the optimal investment, by simply replacing deterministic coefficients by functions, which relate coefficients to an external factor and thus adapt the strategy to an underlying filtration. This is no longer obvious as far as the consumption strategy is concerned. To the best of our knowledge, the formula for the optimal consumption in the model investigated in this paper is new (see Theorem 6.1).

Portfolio optimization in stochastic factor models has recently gained much attention in the financial literature. In the majority of papers, a power utility function is applied and a Black–Scholes financial market with an external stochastic factor of diffusion type is considered. In this setting, it is well known that one must solve a nondegenerate nonlinear second-order partial differential equation. Several methods have been proposed to deal with this problem. In [23], in the case of a pure investment problem, a power transformation was introduced, which makes the nonlinear term disappear. In [16] a similar transformation has been applied, but because of the possible consumption, a linear partial differential equation appears only in the case of perfectly (positively) correlated Brownian motions or for logarithmic utility. More effective methods have been proposed in [12] and [8]. In the first paper, a change of measure transformation is applied and the resulting optimization control problem is investigated, whose value function depends only on time and a factor variable. In the second paper, the dual problem is considered, whose control process belongs to a set of equivalent local martingale measures. Again, the value function of the dual problem depends only on time and a factor variable. This method has also been successfully applied in a robust utility maximization model in [15] recently. In all three aforementioned papers, the existence of a classical solution to the Hamilton–Jacobi–Bellman equation is proved in three steps: first, by constraining the values of the control process to a compact set, second, by applying results from the theory of nondegenerate linear partial differential equations (see Chapter VI.6 and Appendix E in [13]), thus showing that the constrained problem has a unique classical solution, and, third, by studying the asymptotic limit. It seems that this method cannot be successfully applied to our problem.

In the present paper, a candidate solution is first derived heuristically via the Feynman–Kac representation. This leads to a fixed point equation. The existence of a solution is established by Banach's fixed point theorem and its differentiability is proved by using the properties of a suitable operator. Finally, we show that the candidate solution satisfies our integro-differential equation and that this solution is unique. The idea of finding a solution to a control problem through a fixed point theorem is not new; it is, for example, mentioned in [8]. In [4], the existence of a solution to a nondegenerate nonlinear (semilinear) partial differential equation is proved by Banach's fixed point theorem. The smoothness then follows from Hölder estimates for a solution of a nondegenerate linear partial differential equation. We would like to point out that, in particular, in [4], an exponent in the Feynman–Kac formula is assumed to be bounded, which leads to a bounded solution, while we are dealing with a solution which satisfies only an exponential growth condition. We would also like to mention that in the context of optimal control, the results from [4] are directly applied in [10], where an investment and consumption problem is investigated in the presence of default, triggered by a one-jump counting process with a stochastic intensity of diffusion type.

Throughout this paper, we assume that the external factor is observable (as in all aforementioned publications). An alternative would be a partially observed control problem, whose optimal strategy would then be based on an estimate of the underlying factor. We refer to [3] or [21], where a portfolio problem is solved in a diffusion setting with an unobserved volatility process of diffusion type and of Markov switching type, respectively.

Our paper is structured as follows. In Section 2, we introduce the financial market. The optimization problem is formulated in Section 3. The uniqueness of a solution is proved in Section 4, whereas the differentiability is established in Section 5. In Section 6, we show the optimality of a solution and illustrate our findings by means of a numerical example. We also present the solution to the optimal investment and consumption problem for logarithmic utility.

**2. The financial market.** Let $(\Omega, \mathcal{F}, \mathbb{P})$ be a probability space with filtration $\mathbb{F} = (\mathcal{F}(t))_{0 \leq t \leq T}$, where $T$ denotes a finite time horizon. The filtration is assumed to satisfy the usual conditions of completeness and right continuity. The measure $\mathbb{P}$ is the real-world, objective probability measure. All expectations are taken with respect to $\mathbb{P}$.

We consider a Black–Scholes market with coefficients driven by an external stochastic factor. Let $Y := (Y(t))_{0 \leq t \leq T}$ denote this economic factor, whose dynamics is given by a stochastic differential (SDE) equation of the Ornstein–Uhlenbeck type,

$$(2.1) \qquad dY(t) = -\lambda Y(t-)\, dt + dL(\lambda t), \qquad Y(0) = y > 0,$$

where $\lambda > 0$ denotes the reversion rate and $L := (L(t))_{0 \leq t \leq T}$ is an $\mathbb{F}$-adapted subordinator with càdlàg sample paths. Recall that a subordinator is a Lévy process with a.s. nondecreasing sample paths. For definitions and more background on Lévy processes, we refer to [1, 7] or [22].

Our financial market consists of two instruments. The price of a (locally) risk-free asset $B := (B(t))_{0 \leq t \leq T}$ is described by the differential equation

$$(2.2) \qquad \frac{dB(t)}{B(t)} = r(Y(t-))\, dt, \qquad B(0) = 1,$$

whereas the dynamics of the price of a risky asset, $S := (S(t))_{0 \leq t \leq T}$, is given by the SDE

$$(2.3) \qquad \frac{dS(t)}{S(t)} = \mu(Y(t-))\, dt + \sigma(Y(t-))\, dW(t), \qquad S(0) = s > 0,$$

where $W := (W(t))_{0 \leq t \leq T}$ denotes an $\mathbb{F}$-adapted Brownian motion, independent of the subordinator $L$. We make the following assumptions concerning the functions $r, \mu$ and $\sigma$:

(A1) the functions $r: (0, \infty) \to [0, \infty)$, $\mu: (0, \infty) \to [0, \infty)$ and $\sigma: (0, \infty) \to (0, \infty)$ are continuous and satisfy the linear growth conditions

$$r(y) \leq A_r + B_r y, \qquad \mu(y) \leq A_\mu + B_\mu y,$$
$$\sigma^2(y) \leq A_\sigma + B_\sigma y, \qquad y > 0,$$

with nonnegative constants;

(A2) the derivatives $\frac{dr}{dy}:(0,\infty)\to\mathbb{R}, \frac{d\mu}{dy}:(0,\infty)\to\mathbb{R}$ and $\frac{d\sigma^2}{dy}:(0,\infty)\to\mathbb{R}$ are continuous and satisfy linear growth conditions analogous to those of $r, \mu, \sigma^2$;

(A3) $\inf_{y\in\mathcal{D}_2}\sigma(y) > 0$, where the set $\mathcal{D}_2$ will be specified in (3.7).

Note that the assumptions (A1)–(A3) are more general than in [8, 12] and [15], where uniform boundedness of the functions $r, \mu, \sigma^2$ and their first derivatives is required. Our conditions are similar to those in [23], where Lipschitz continuity of the coefficients is assumed, together with a linear growth condition.

A prominent example of the above financial model is the Barndorff-Nielsen and Shephard model, introduced in [6], which can be described by the following set of equations:

$$\frac{dB(t)}{B(t)} = r\,dt,$$

(2.4) $$\frac{dS(t)}{S(t)} = (\mu + \beta Y(t-))\,dt + \sqrt{Y(t-)}\,dW(t).$$

Besides the above paper, we also refer to [9], Chapter 15, [5, 17] and references therein for more information about the properties of non-Gaussian stochastic volatility models in the context relevant to our paper.

We shall need some further results and notation for $Y$ and its background driving Lévy process $L$. The subordinator $L$ has the representation (see, e.g., [1], Chapter 1.3.2)

(2.5) $$L(t) = \int_{(0,t]}\int_{z>0} z N(ds,dz), \qquad t \geq 0,$$

where $N((0,t]\times A) = \#\{0 < s \leq t : (L(s) - L(s-)) \in A\}$ denotes a Poisson random measure with a deterministic, time-homogeneous intensity measure $\nu(dz)\,ds$ satisfying $\int_{0<z<1} z\nu(dz) < \infty$. The fundamental result in the theory of infinitely divisible random variables is the Lévy–Kintchine formula, which presents the moment generating function of a subordinator as

(2.6) $$\mathbb{E}[e^{wL(t)}] = e^{t\psi(w)} = \exp\left\{t\int_{z>0}(e^{wz}-1)\nu(dz)\right\}, \qquad w \leq \bar{w},$$

for some $\bar{w} \in [0,\infty]$. The function $\psi(w)$ is called the Laplace exponent of $L$. Note that $\psi(w)$ exists at least for all $w \leq 0$ and $\psi(w) > 0$ for $w > 0$, provided it exists.

Let us now investigate the SDE (2.1). Its unique solution for $s > t$ is given by (cf. [1], Chapter 6.3)

(2.7) $$Y(s) = ye^{-\lambda(s-t)} + \int_t^s e^{-\lambda(u-t)}\,dL(\lambda u), \qquad Y(t) = y.$$

We abbreviate the process (2.7) by $Y^{t,y} := (Y^{t,y}(s), t \leq s \leq T)$ and would like to point out that it has a.s. càdlàg sample paths of finite variation and that the mapping $y \mapsto Y^{t,y}$ is continuous $\mathbb{P}$-a.s. Moreover,

$$\frac{\partial}{\partial y} Y^{t,y}(s) = e^{-\lambda(s-t)}, \qquad \mathbb{P}\text{-a.s.} \tag{2.8}$$

Finally, it is straightforward to establish the following relations for all $0 \leq t \leq s \leq T$ and $y > 0$:

$$Y^{t,y}(s) \leq y + L(\lambda s) - L(\lambda t), \tag{2.9}$$

$$\lambda \int_t^s Y^{t,y}(u)\,du = y + L(\lambda s) - L(\lambda t) - Y^{t,y}(s) \tag{2.10}$$

$$\leq y + L(\lambda s) - L(\lambda t) = y + L(\lambda(s-t)). \tag{2.11}$$

The above relations hold $\mathbb{P}$-a.s., except for the last equality, which holds in distribution.

**3. Formulation of the optimization problem.** We consider an investor who makes decisions concerning investment and consumption of a portfolio based on a power utility function of the form $x^\gamma$ for $\gamma \in (0,1)$.

Consider the wealth process $X^{c,\pi} := (X^{c,\pi}(t))_{0 \leq t \leq T}$ of an agent. Its dynamics is given by the stochastic differential equation

$$\begin{aligned} dX^{c,\pi}(t) &= \pi(t) X^{c,\pi}(t)(\mu(Y(t-))\,dt + \sigma(Y(t-))\,dW(t)) \\ &\quad + (1-\pi(t)) X^{c,\pi}(t) r(Y(t-))\,dt - c(t)\,dt, \end{aligned} \tag{3.1}$$

where $\pi(t)$ denotes a fraction of the wealth invested in the risky asset and $c(t)$ denotes the rate of consumption at time $t$. We are dealing with the following optimization problem:

$$\sup_{c,\pi} \mathbb{E}\left[\int_0^T (c(s))^\gamma\,ds + (X^{c,\pi}(T))^\gamma \mid X(0) = x, Y(0) = y\right]. \tag{3.2}$$

The corresponding optimal value function is defined as

$$V(t,x,y) = \sup_{(c,\pi) \in \mathcal{A}} \mathbb{E}\left[\int_t^T (c(s))^\gamma\,ds + (X^{c,\pi}(T))^\gamma \mid X(t) = x, \right.$$

$$\left. Y(t) = y\right]. \tag{3.3}$$

Let us introduce the set $\mathcal{A}$ of admissible strategies.

DEFINITION 3.1. A strategy $(c,\pi) := (c(t), \pi(t))_{0 \leq t \leq T}$ is *admissible*, and we write $(c,\pi) \in \mathcal{A}$, if it satisfies the following conditions:

1. $(c,\pi)\colon (0,T]\times\Omega\to [0,\infty)\times [0,1]$ is a progressively measurable mapping with respect to the filtration $\mathbb{F}$;
2. $\int_0^T c(s)\,ds < \infty$ $\mathbb{P}$-a.s.;
3. the SDE (3.1) has a unique, positive solution $X^{c,\pi}$ on $[0,T]$.

We would like to mention that for every $(c,\pi)\in\mathcal{A}$, the wealth process $X^{c,\pi}$, which satisfies (3.1), is an Itô diffusion; that is, in particular, a semi-martingale with $\mathbb{P}$-a.s. continuous sample paths.

Note that we exclude the possibility of borrowing from the bank account and short-selling the asset, as in [5] and [17]. Technically, there is no problem in solving the unconstrained optimization problem. In particular, if $(\mu(y)-r(y))/\sigma^2(y)$ is positive and uniformly bounded, then all of our results remain the same.

One can associate a Hamilton–Jacobi–Bellman equation with the optimization problem (3.3) given by the following partial integro-differential equation

$$
\begin{aligned}
(3.4)\quad \sup_{(c,\pi)\in[0,\infty)\times[0,1]} \Bigg\{ & c^\gamma + \frac{\partial v}{\partial t}(t,x,y) \\
& + \frac{\partial v}{\partial x}(t,x,y)(\pi x(\mu(y)-r(y)) + xr(y) - c) \\
& + \frac{1}{2}\frac{\partial^2 v}{\partial x^2}(t,x,y)\pi^2 x^2 \sigma^2(y) - \frac{\partial v}{\partial y}(t,x,y)\lambda y \\
& + \lambda \int_{z>0} (v(t,x,y+z) - v(t,x,y))\nu(dz) \Bigg\} = 0, \\
& v(T,x,y) = x^\gamma.
\end{aligned}
$$

As we use a power utility function, it is natural to try to find a solution of the form $v(t,x,y) = x^\gamma f(t,y)$ for some function $f$. With this choice of value function, the optimal strategy $(\hat{c},\hat{\pi})$, which maximizes the left-hand side of (3.4), is given by

$$(3.5)\qquad \hat{c} = xf(t,y)^{-1/(1-\gamma)},$$

$$(3.6)\qquad \hat{\pi} = \arg\max_{\pi\in[0,1]}\{\pi(\mu(y)-r(y)) - \tfrac{1}{2}\pi^2(1-\gamma)\sigma^2(y)\}.$$

To investigate the formula for the investment strategy more closely, we define the three sets

$$
\begin{aligned}
\mathcal{D}_1 &= \{y>0, \mu(y)-r(y)<0\}, \\
(3.7)\quad \mathcal{D}_2 &= \{y>0, \mu(y)-r(y)>0, (1-\gamma)\sigma^2(y) > \mu(y)-r(y)\}, \\
\mathcal{D}_3 &= \{y>0, \mu(y)-r(y)>0, (1-\gamma)\sigma^2(y) < \mu(y)-r(y)\}.
\end{aligned}
$$

The strategy $\hat{\pi}$ is given by

$$
(3.8) \qquad \hat{\pi} = \begin{cases} 0, & y \in \mathcal{D}_1, \\ \dfrac{\mu(y) - r(y)}{(1-\gamma)\sigma^2(y)}, & y \in \mathcal{D}_2, \\ 1, & y \in \mathcal{D}_3. \end{cases}
$$

The following lemma is a counterpart of Lemma 5.1 in [5].

LEMMA 3.2. *Define the function*

$$
(3.9) \qquad Q(y) = \max_{\pi \in [0,1]} \left\{ \pi(\mu(y) - r(y)) - \frac{1}{2}\pi^2(1-\gamma)\sigma^2(y) \right\} + r(y)
$$

$$
= \begin{cases} r(y), & y \in \mathcal{D}_1, \\ \dfrac{(\mu(y) - r(y))^2}{2(1-\gamma)\sigma^2(y)} + r(y), & y \in \mathcal{D}_2, \\ \mu(y) - \dfrac{1}{2}(1-\gamma)\sigma^2(y), & y \in \mathcal{D}_3. \end{cases}
$$

*The function $Q$ is nonnegative, continuous and satisfies the linear growth condition*

$$
(3.10) \qquad 0 \le r(y) \le Q(y) \le A + By, \qquad y > 0,
$$

*for nonnegative $A$ and $B$. The derivative of $Q$ is continuous and also satisfies a linear growth condition: for nonnegative $C$ and $D$, we have*

$$
\left| \frac{dQ}{dy}(y) \right| \le C + Dy, \qquad y > 0.
$$

PROOF. First, note that the sets $\mathcal{D}_1$ and $\mathcal{D}_2$ have common boundary

$$
(3.11) \qquad \partial \mathcal{D}_{12} = \{ y > 0, \mu(y) = r(y) \}
$$

and that $\mathcal{D}_2$ and $\mathcal{D}_3$ have common boundary

$$
(3.12) \qquad \partial \mathcal{D}_{23} = \{ y > 0, (1-\gamma)\sigma^2(y) = \mu(y) - r(y) \}.
$$

The sets $\mathcal{D}_1$ and $\mathcal{D}_3$ do not have a common boundary.

It is straightforward to show that $Q$ is continuous in $\mathcal{D}_1$, $\mathcal{D}_2$ and $\mathcal{D}_3$, as well as over the boundaries $\partial \mathcal{D}_{12}$ and $\partial \mathcal{D}_{23}$. The linear growth condition clearly holds in the sets $\mathcal{D}_1$ and $\mathcal{D}_3$. Note that in $\mathcal{D}_2$, the inequality

$$
(3.13) \qquad \begin{aligned} \frac{(\mu(y) - r(y))^2}{2(1-\gamma)\sigma^2(y)} + r(y) &\le \frac{1}{2}(\mu(y) - r(y)) + r(y) \\ &= \frac{1}{2}(\mu(y) + r(y)) \end{aligned}
$$

holds, from which the linear growth condition of the function $Q$ in the set $\mathcal{D}_2$ follows, from (A1).

We differentiate the function $Q$ and obtain

$$\frac{dQ}{dy}(y)$$
$$= \begin{cases} \frac{dr}{dy}(y), \\ \qquad y \in \mathcal{D}_1, \\ \frac{(\mu(y) - r(y))(\frac{d\mu}{dy}(y) - \frac{dr}{dy}(y))}{(1-\gamma)\sigma^2(y)} - \frac{(\mu(y) - r(y))^2 \frac{d\sigma}{dy}(y)}{(1-\gamma)\sigma^3(y)} + \frac{dr}{dy}(y), \\ \qquad y \in \mathcal{D}_2, \\ \frac{d\mu}{dy}(y) - \frac{1}{2}(1-\gamma)\frac{d\sigma^2}{dy}(y), \qquad y \in \mathcal{D}_3. \end{cases}$$

Again, it is easy to show that this derivative is continuous in $\mathcal{D}_1$, $\mathcal{D}_2$ and $\mathcal{D}_3$ and over the boundaries $\partial \mathcal{D}_{12}$ and $\partial \mathcal{D}_{23}$, and that a linear growth condition holds in $\mathcal{D}_1$ and $\mathcal{D}_3$. To prove the linear growth condition in the set $\mathcal{D}_2$, note that

$$\left|\frac{dQ}{dy}(y)\right| \leq \left|\frac{d\mu}{dy}(y)\right| + \frac{1}{2}(1-\gamma)\left|\frac{d\sigma^2}{dy}(y)\right|$$

holds for $y \in \mathcal{D}_2$. $\square$

REMARK 3.3. When investigating the unconstrained optimization problem, $\pi \in \mathbb{R}$, the set $\mathcal{D}_2$ must coincide with the whole positive real line and one must assume a uniform lower bound of the function $\sigma$, that is, $\inf_{y>0} \sigma(y) > 0$. In the Barndorff-Nielsen and Shephard model, this condition does not hold unless we introduce reversion to a strictly positive constant (i.e., a linear drift term with positive mean reverting level). However, by considering a constrained strategy, one can overcome the global lower uniform boundedness and work with uniform boundedness only over some subset; see [5] and [17] for the structure of the set $\mathcal{D}_2$. Note that for the constrained optimization problem, condition (A3) is not necessary. If volatility hits zero, one can assume that the set $\mathcal{D}_2$ reduces to an empty set so that the results from this paper remain valid. However, in order that all terms in (3.8) and (3.9) are well defined, we prefer to retain condition (A3). Moreover, we point out that (A3) is very common in stochastic volatility optimization models (see [8, 12, 15, 23]) as well as being economically sensible.

We would like to point out that the linear growth condition (3.10) and the relations (2.8)–(2.11) will be frequently applied when proving our results.

By substituting (3.5) and (3.6) into (3.4) we arrive at the nonlinear first-order partial integro-differential equation for the function $f$,

$$\begin{aligned}
0 &= \frac{\partial f}{\partial t}(t,y) - \lambda \frac{\partial f}{\partial y}(t,y)y + \lambda \int_{z>0} (f(t, y+z) - f(t,y))\nu(dz) \\
&\quad + \gamma f(t,y)Q(y) + (1-\gamma)f(t,y)^{-\gamma/(1-\gamma)}, \qquad f(T,y) = 1.
\end{aligned} \tag{3.14}$$

We will show that there exists a unique classical solution to this equation.

**4. Existence of the solution.** We introduce an operator $\mathcal{L}$ acting on functions $f$ as follows:

$$(\mathcal{L}f)(t,y) = \mathbb{E}\bigg[ e^{\gamma \int_t^T Q(Y^{t,y}(s))ds} \\
+ (1-\gamma)\int_t^T e^{\gamma \int_t^s Q(Y^{t,y}(u))du} f(s, Y^{t,y}(s))^{-\gamma/(1-\gamma)}\, ds \bigg], \tag{4.1}$$

for $Q$ as in Lemma 3.2. By applying (heuristically) the Feynman–Kac formula to (3.14), we arrive at the following fixed point equation:

$$(\mathcal{L}f)(t,y) = f(t,y), \qquad (t,y) \in [0,T] \times (0,\infty). \tag{4.2}$$

In this section, we prove that equation (4.2) has a unique solution $\hat{f}$. In Section 5, we shall show that this solution satisfies the partial integro-differential equation (3.14) in the classical sense.

We start with some observations. Note that it is easy to derive a lower bound for the optimal value function $V$,

$$V(t,x,y) \geq x^\gamma \mathbb{E}[e^{\gamma \int_t^T r(Y(s))ds}], \tag{4.3}$$

$$(t,x,y) \in [0,T] \times (0,\infty) \times (0,\infty),$$

where the left-hand side of (4.3) is the payoff, when the agent does not consume and invests everything in the bank account. We conclude that the solution to (4.2) should satisfy the inequality

$$f(t,y) \geq \mathbb{E}[e^{\gamma \int_t^T r(Y(s))ds}] \geq 1, \qquad (t,y) \in [0,T] \times (0,\infty). \tag{4.4}$$

Moreover, if we apply the operator $\mathcal{L}$ to a function $f$ which satisfies (4.4), then we obtain a lower bound of the operator,

$$(\mathcal{L}f)(t,y) \geq \mathbb{E}[e^{\gamma \int_t^T r(Y(s))ds}] \geq 1, \qquad (t,y) \in [0,T] \times (0,\infty), \tag{4.5}$$

which can be derived by noting that the second term in (4.1) is positive and by applying the lower estimate (3.10) of the function $Q$ in the first term.

We now turn to the more interesting upper bound of the operator $\mathcal{L}$. We still assume that condition (4.4) holds, which implies that $f(t,y)^{-\gamma/(1-\gamma)} \leq 1$.

By applying the upper estimate (3.10) of the function $Q$, the estimate (2.11) and the representation (2.6), provided that $\psi(\gamma B/\lambda) < \infty$, we obtain the inequality

$$
\begin{aligned}
(\mathcal{L}f)(t,y) &\leq \mathbb{E}\bigg[e^{\gamma A(T-t)+\gamma B \int_t^T Y^{t,y}(s)ds} \\
&\qquad + (1-\gamma)\int_t^T e^{\gamma A(s-t)+\gamma B \int_t^s Y^{t,y}(u)du}\,ds\bigg] \\
&\leq \mathbb{E}\bigg[e^{\gamma A(T-t)+(\gamma B/\lambda)y+(\gamma B/\lambda)(L(\lambda T)-L(\lambda t))} \\
&\qquad + (1-\gamma)\int_t^T e^{\gamma A(s-t)+(\gamma B/\lambda)y+(\gamma B/\lambda)(L(\lambda s)-L(\lambda t))}\,ds\bigg] \\
&= e^{\gamma A(T-t)+(\gamma B/\lambda)y+\lambda\psi(\gamma B/\lambda)(T-t)} \\
&\qquad + (1-\gamma)\int_t^T e^{\gamma A(s-t)+(\gamma B/\lambda)y+\lambda\psi(\gamma B/\lambda)(s-t)}\,ds \\
&\leq \left(1+\frac{1-\gamma}{A'}\right)e^{A'(T-t)+B'y},
\end{aligned}
\tag{4.6}
$$

where we have introduced the constants $A' = \gamma A + \lambda\psi(\gamma B/\lambda) > 0$ and $B' = \gamma B/\lambda \geq 0$.

In the rest of the paper, we assume that the following condition on the Lévy measure of $L$, formulated in terms of the characteristic exponent in (2.6) holds:

(B) $\psi(w) < \infty$ for $w = 2(1+\frac{\gamma}{2})(B' \vee B'_\sigma) + \varepsilon$ and some $\varepsilon > 0$,

where $B'_\sigma = \gamma B_\sigma/\lambda \geq 0$ is defined analogously to $B'$ and $B_\sigma$ is defined in (A1). The reason for this assumption becomes clear in the course of our calculations. It is needed in Section 6 in order to verify the optimality. Note that the lemmas in this section and Section 5 hold under integrability conditions of lower orders.

Let us investigate the operator $\mathcal{L}$ in a more rigorous way. Denote by $\mathcal{C}_e([0,T] \times (0,\infty))$ the space of continuous functions $f$ on $[0,T] \times (0,\infty)$ satisfying

$$1 \leq f(t,y) \leq \left(1 + \frac{1-\gamma}{A'}\right)e^{A'(T-t)+B'y}, \qquad (t,y) \in [0,T] \times (0,\infty).$$

We define a metric an $\mathcal{C}_e([0,T] \times (0,\infty))$ by

$$\mathrm{d}(\varphi,\xi) = \sup_{(t,y)\in[0,T]\times(0,\infty)} |e^{-\alpha(T-t)-B'y}(\varphi(t,y)-\xi(t,y))|, \tag{4.7}$$

for some $\alpha > A'$ to be specified later. The space $(\mathcal{C}_e([0,T] \times (0,\infty)), d)$ is a complete metric space. Below, we state two lemmas dealing with the properties of the operator $\mathcal{L}$.

LEMMA 4.1. *The operator $\mathcal{L}$ defines a mapping from $\mathcal{C}_e([0,T] \times (0,\infty))$ into itself.*

PROOF. Based on our previous results (4.5) and (4.6), we can conclude that the lower and upper bounds are preserved. It remains to prove the continuity of the mapping $(t,y) \mapsto (\mathcal{L}f)(t,y)$. Due to the time homogeneity of $Y$, the operator $\mathcal{L}$ can be represented as

$$(\mathcal{L}f)(t,y) = \mathbb{E}\bigg[e^{\gamma \int_0^{T-t} Q(Y^{0,y}(s))ds}$$
$$+ (1-\gamma) \int_0^{T-t} e^{\gamma \int_0^s Q(Y^{0,y}(u))du} f(s+t, Y^{0,y}(s))^{-\gamma/(1-\gamma)}\, ds\bigg].$$

The above representation simplifies proving continuity in the time variable. Note that by the growth condition (3.10) and relation (2.11),

$$e^{\gamma \int_0^s Q(Y^{0,y}(u))du} f(s+t, Y^{0,y}(s))^{-\gamma/(1-\gamma)} \leq e^{\gamma \int_0^s Q(Y^{0,y}(s))ds}$$
$$\leq e^{\gamma AT + B'y + B'L(\lambda T)}$$

holds $\mathbb{P}$-a.s. and the càdlàg mapping $(y,u) \mapsto Y^{0,y}(u)$ is bounded a.s on compact sets. In order to prove continuity in the time variable, one can directly apply Lebesgue's dominated convergence theorem and take the limit under the integral. To prove continuity of the mapping $y \mapsto (\mathcal{L}f)(t,y)$ at a fixed point $y_0 > 0$, define a compact set $U$ around $y_0$ and take a sequence of points $y_n \in U$ such that $y_n \to y_0$ as $n \to \infty$. In this setting, we can find a uniform bound for all $y_n \in U$ and can apply Lebesgue's dominated convergence theorem. The continuity of $(t,y) \mapsto (\mathcal{L}f)(t,y)$ now follows from the continuity of $f$ and $Q$ and the continuity of the mapping $y \mapsto Y^{0,y}$. □

LEMMA 4.2. *The mapping $\mathcal{L}: \mathcal{C}_e([0,T] \times (0,\infty)) \to \mathcal{C}_e([0,T] \times (0,\infty))$ is a contraction with respect to the metric (4.7) for $\alpha > A' + \gamma$.*

PROOF. Take two functions $\varphi, \xi \in \mathcal{C}_e([0,T] \times (0,\infty))$. Again, we invoke (2.10) and (3.10). The following inequalities then hold for all $(t,y) \in [0,T] \times (0,\infty)$:

$$\mathrm{d}(\mathcal{L}\varphi, \mathcal{L}\xi) = |e^{-\alpha(T-t)-B'y}((\mathcal{L}\varphi)(t,y) - (\mathcal{L}\xi)(t,y))|$$
$$\leq (1-\gamma)e^{-\alpha(T-t)-B'y}$$
$$\times \mathbb{E}\bigg[\int_t^T e^{r\gamma \int_t^s Q(Y^{t,y}(u))du} |\varphi(s, Y^{t,y}(s))^{-\gamma/(1-\gamma)}$$
$$- \xi(s, Y^{t,y}(s))^{-\gamma/(1-\gamma)}|\, ds\bigg]$$

$$\leq \gamma e^{-\alpha(T-t)-B'y}\mathbb{E}\left[\int_t^T e^{\gamma \int_t^s Q(Y^{t,y}(u))du}|\varphi(s,Y^{t,y}(s))\right.$$
$$\left. - \xi(s,Y^{t,y}(s))|\,ds\right]$$
$$\leq \gamma e^{-\alpha(T-t)-B'y}\mathrm{d}(\varphi,\xi)\mathbb{E}\left[\int_t^T e^{\gamma \int_t^s Q(Y^{t,y}(u))du+\alpha(T-s)+B'Y^{t,y}(s)}\,ds\right]$$
$$\leq \gamma e^{-\alpha(T-t)-B'y}\mathrm{d}(\varphi,\xi)$$
$$\times \mathbb{E}\left[\int_t^T e^{\gamma A(s-t)+B'(y+L(\lambda s)-L(\lambda t)-Y^{t,y}(s))+\alpha(T-s)+B'Y^{t,y}(s)}\,ds\right]$$
$$= \gamma \mathrm{d}(\varphi,\xi)\int_t^T e^{-\alpha(s-t)+\gamma A(s-t)+\lambda\psi(B')(s-t)}\,ds$$
$$\leq \frac{\gamma}{\alpha - A'}\mathrm{d}(\varphi,\xi),$$

where the mean value theorem has been applied in line 3. We conclude that

$$\mathrm{d}(\mathcal{L}\phi,\mathcal{L}\xi) \leq \zeta \mathrm{d}(\varphi,\xi), \qquad \zeta < 1,$$

which proves that the operator $\mathcal{L}$ defines a contraction mapping. $\square$

The main result of this section is the following proposition, which is a consequence of Banach's fixed point theorem.

PROPOSITION 4.3. *The equation*

(4.8) $$(\mathcal{L}f)(t,y) = f(t,y)$$

*has a unique solution* $\hat{f} \in \mathcal{C}_e([0,T] \times (0,\infty))$.

**5. Differentiability of the solution.** In this section, we establish the differentiability of the function $\hat{f}$. In order to apply a classical verification theorem, we have to prove that $\hat{f}$ is continuously differentiable in the time and in the space variable.

We assume that $B > 0$. In the case when the function $Q$ is uniformly bounded in $y$, that is, $B = 0$, an arbitrary, strictly positive (small) constant $B > 0$ can be chosen so that the proofs from this section remain true. We remark that our arguments can be modified in order to handle the special case of $B = 0$ and to derive sharper bounds. We would like to point out that the main theorem of our paper, Theorem 6.1, holds true even for $B = 0$.

Recall that the ODE

(5.1) $$\frac{d\phi}{dt}(t) + (\gamma - \lambda)\phi(t) + \lambda a = 0, \qquad \phi(T) = a,$$

has the unique, smooth and strictly positive solution in the class $\mathcal{C}^1([0,T])$ given by

$$\phi(t) = a + \gamma \int_t^T \phi(s) e^{-\lambda(s-t)} \, ds,$$

with constant $a > 0$.

The idea for establishing differentiability in the space variable is to construct a sequence of functions $(f_n)_{n \in \mathbb{N}}$ which converge to $\hat{f}$ and which share some desirable properties.

LEMMA 5.1. *Define* $A'' = \gamma A + \lambda \psi(B'') > 0$ *and* $B'' = B'(1 + \frac{\gamma}{4}) > 0$, *with* $B' = \gamma B/\lambda > 0$ *and* $A, B$ *as in (3.10).*
*Choose a function* $f_1 \in \mathcal{C}_e([0,T] \times (0,\infty)) \cap \mathcal{C}^{0,1}([0,T] \times (0,\infty))$ *such that*

(5.2) $$\left| \frac{\partial f_1}{\partial y}(t,y) \right| \leq \phi(t) e^{A''(T-t) + B''y}, \qquad (t,y) \in [0,T] \times (0,\infty),$$

*where* $\phi$ *solves (5.1) with* $a = \frac{1}{\lambda}(1 + \frac{1-\gamma}{A''})(\frac{4D}{B'} \vee C\gamma) > 0$ *and* $C, D$ *as in Lemma 3.2.*

*Now, construct now the sequence* $(f_n)_{n \in \mathbb{N}}$ *recursively as* $f_{n+1} = \mathcal{L} f_n$ *with* $\mathcal{L}$ *defined as in (4.1).*

*Then, for all* $n \in \mathbb{N}$,

$$f_n \in \mathcal{C}_e([0,T] \times (0,\infty)) \cap \mathcal{C}^{0,1}([0,T] \times (0,\infty))$$

*and*

(5.3) $$\left| \frac{\partial f_n}{\partial y}(t,y) \right| \leq \phi(t) e^{A''(T-t) + B''y}, \qquad (t,y) \in [0,T] \times (0,\infty).$$

PROOF. Recall from (4.1) that

$$f_2(t,y) = \mathbb{E}\bigg[ e^{\gamma \int_t^T Q(Y^{t,y}(s)) ds}$$
$$+ (1-\gamma) \int_t^T e^{\gamma \int_t^s Q(Y^{t,y}(u)) du} f_1(s, Y^{t,y}(s))^{-\gamma/(1-\gamma)} \, ds \bigg].$$

First, we prove that the mapping $(t,y) \mapsto \frac{\partial f_2}{\partial y}(t,y)$ is continuous.

We expect that the derivative equals

$$\frac{\partial f_2}{\partial y}(t,y) = \mathbb{E}\bigg[ \gamma e^{\gamma \int_t^T Q(Y^{t,y}(s)) ds} \int_t^T \frac{dQ}{dy}(Y^{t,y}(s)) e^{-\lambda(s-t)} \, ds$$
$$+ (1-\gamma) \int_t^T \gamma e^{\gamma \int_t^s Q(Y^{t,y}(u)) du} f_1(s, Y(s))^{-\gamma/(1-\gamma)}$$
(5.4) $$\times \int_t^s \frac{dQ}{dy}(Y^{t,y}(u)) e^{-\lambda(u-t)} \, du \, ds$$

$$- \int_t^T \gamma e^{\gamma \int_t^s Q(Y^{t,y}(u))du} f_1(s, Y^{t,y}(s))^{-1/(1-\gamma)}$$
$$\times \frac{\partial f_1}{\partial y}(s, Y^{t,y}(s)) e^{-\lambda(s-t)} \, ds \bigg].$$

This will follow from Lebesgue's dominated convergence theorem, provided that it can be applied. Below, we establish three estimates which allow us to interchange differentiation and integration. We point out that the interchange is justified if we can bound the derivative by an integrable function. The estimates are also used later to establish (5.3).

We recall that in order to find a uniform bound, one can take a limit $y_n \to y_0$, as $n \to \infty$, over a sequence of points $y_n \in U$, where $U$ is a compact set around a fixed point $y_0 > 0$.

Note that by invoking the simple inequality $a + by \leq (\frac{1}{\epsilon} \vee a) e^{b\epsilon y}$ for all $\epsilon > 0$, together with (2.9), we obtain that

(5.5)
$$\left| \frac{\partial}{\partial y}(Q(Y^{t,y}(u))) \right| = \left| \frac{dQ}{dy}(Y^{t,y}(u)) \frac{\partial}{\partial y} Y^{t,y}(u) \right|$$
$$\leq (C + D Y^{t,y}(u)) e^{-\lambda(u-t)}$$
$$\leq \left( \frac{4D}{\gamma B'} \vee C \right) e^{(\gamma B'/4) Y^{t,y}(u)} e^{-\lambda(u-t)}$$
$$\leq \left( \frac{4D}{\gamma B'} \vee C \right) e^{(\gamma B'/4)(y+L(\lambda s)-L(\lambda t))} e^{-\lambda(u-t)}$$

holds $\mathbb{P}$-a.s., for $0 \leq t \leq u \leq s \leq T$. Based on (5.5), we derive that

(5.6)
$$\left| \frac{\partial}{\partial y}(e^{\gamma \int_t^s Q(Y^{t,y}(u))du}) \right|$$
$$= \left| \gamma e^{\gamma \int_t^s Q(Y^{t,y}(u))du} \frac{\partial}{\partial y}\left( \int_t^s Q(Y^{t,y}(u)) \, du \right) \right|$$
$$= \left| \gamma e^{\gamma \int_t^s Q(Y^{t,y})(u)du} \int_t^s \frac{\partial}{\partial y}(Q(Y^{t,y}(u))) \, du \right|$$
$$\leq \left( \frac{4D}{B'} \vee C\gamma \right)$$
$$\times e^{\gamma A(s-t) + B'(y+L(\lambda s)-L(\lambda t))} e^{(\gamma B'/4)(y+L(\lambda s)-L(\lambda t))} \int_t^s e^{-\lambda(u-t)} \, du$$
$$\leq \frac{1}{\lambda}\left( \frac{4D}{B'} \vee C\gamma \right) e^{\gamma A(s-t) + B''(y+L(\lambda s)-L(\lambda t))}$$

holds $\mathbb{P}$-a.s., for $0 \leq t \leq s \leq T$. We would like to point out that we are allowed to interchange integration and differentiation in the first line of (5.6) since the bound (5.5) is integrable $\mathbb{P}$-a.s. on $[t, s]$.

Based on (5.2) and (5.6), we obtain the third estimate

$$\left| \frac{\partial}{\partial y}((1-\gamma)e^{\gamma \int_t^s Q(Y^{t,y}(u))du} f_1(s,Y(s))^{-\gamma/(1-\gamma)}) \right|$$

$$= \left| (1-\gamma)\frac{\partial}{\partial y}(e^{\gamma \int_t^s Q(Y^{t,y}(u))du}) f_1(s,Y(s))^{-\gamma/(1-\gamma)} \right.$$

$$\left. - \gamma e^{\gamma \int_t^s Q(Y^{t,y}(u))du} f_1(s,Y^{t,y}(s))^{-1/(1-\gamma)} \frac{\partial f_1}{\partial y}(s,Y^{t,y}(s))\frac{\partial}{\partial y}(Y^{t,y}(s)) \right|$$

$$\leq \frac{(1-\gamma)}{\lambda}\left(\frac{4D}{B'} \vee C\gamma\right) e^{\gamma A(s-t)+B''(y+L(\lambda s)-L(\lambda t))}$$

(5.7)     $+ \gamma e^{\gamma A(s-t)+\gamma B \int_t^s Y^{t,y}(u)du} \left| \frac{\partial f_1}{\partial y}(s,Y^{t,y}(s))\right| e^{-\lambda(s-t)}$

$$\leq \frac{(1-\gamma)}{\lambda}\left(\frac{4D}{B'} \vee C\gamma\right) e^{\gamma A(s-t)+B''(y+L(\lambda s)-L(\lambda t))}$$

$$+ \gamma e^{\gamma A(s-t)+B''(y+L(\lambda s)-L(\lambda t)-Y^{t,y}(s))} \phi(s) e^{A''(T-s)+B''Y^{t,y}(s)} e^{-\lambda(s-t)}$$

$$= \frac{1-\gamma}{\lambda}\left(\frac{4D}{B'} \vee C\gamma\right) e^{\gamma A(s-t)+B''(y+L(\lambda s)-L(\lambda t))}$$

$$+ \gamma \phi(s) e^{-\lambda(s-t)} e^{A''(T-s)+\gamma A(s-t)+B''y+B''(L(\lambda s)-L(\lambda t))}, \qquad \mathbb{P}\text{-a.s.}$$

As the derived bound (5.7) is a càdlàg mapping, it is a.s. integrable, and we have that

$$\frac{\partial}{\partial y}\int_t^T ((1-\gamma)e^{\gamma \int_t^s Q(Y^{t,y}(u))du} f_1(s,Y(s))^{-\gamma/(1-\gamma)})\,ds$$

$$= \int_t^T \frac{\partial}{\partial y}((1-\gamma)e^{\gamma \int_t^s Q(Y^{t,y}(u))du} f_1(s,Y(s))^{-\gamma/(1-\gamma)})\,ds, \qquad \mathbb{P}\text{-a.s.}$$

Finally, taking the derivative under the expectation is also justified since, by condition (B), we have $\psi(B'') < \infty$. Consequently, we have shown that the derivative (5.4) holds.

The continuity of the mapping $(t,y) \mapsto \frac{\partial f_2}{\partial y}(t,y)$ again follows from Lebesgue's dominated convergence theorem, by applying the estimates (5.6) and (5.7), and from the continuity of the functions $f_1$ and $Q$, as well as their derivatives [cf. the proof of continuity in Lemma (4.1)].

We still have to prove that the bound (5.3) holds. By combining (5.6) and (5.7), we can estimate for $n=2$:

$$\left|\frac{\partial f_2}{\partial y}(t,y)\right|$$

$$\leq \mathbb{E}\left[\frac{1}{\lambda}\left(\frac{4D}{B'} \vee C\gamma\right) e^{\gamma A(T-t)+B''(y+L(\lambda T)-L(\lambda t))}\right.$$

$$+ \frac{1-\gamma}{\lambda}\left(\frac{4D}{B'} \vee C\gamma\right) \int_t^T e^{\gamma A(s-t)+B''(y+L(\lambda s)-L(\lambda t))} \, ds$$

$$+ \gamma \int_t^T \phi(s) e^{-\lambda(s-t)} e^{A''(T-s)+\gamma A(s-t)+B''(y+L(\lambda s)-L(\lambda t))} \, ds \bigg]$$

$$= \frac{1}{\lambda}\left(\frac{4D}{B'} \vee C\gamma\right) e^{A''(T-t)+B''y}$$

$$+ \frac{1-\gamma}{\lambda}\left(\frac{4D}{B'} \vee C\gamma\right) \int_t^T e^{A''(s-t)+B''y} \, ds$$

$$+ e^{A''(T-t)+B''y} \gamma \int_t^T \phi(s) e^{-\lambda(s-t)} \, ds$$

$$\leq \frac{1}{\lambda}\left(\frac{4D}{B'} \vee C\gamma\right) e^{A''(T-t)+B''y}$$

$$+ \frac{1-\gamma}{\lambda A''}\left(\frac{4D}{B'} \vee C\gamma\right) e^{A''(T-t)+B''y}$$

$$+ e^{A''(T-t)+B''y} \gamma \int_t^T \phi(s) e^{-\lambda(s-t)} \, ds$$

$$= e^{A''(T-t)+B''y}\left(a + \gamma \int_t^T \phi(s) e^{-\lambda(s-t)} \, ds\right) = \phi(t) e^{A''(T-t)+B''y},$$

where we have invoked the solution of the ODE (5.1) with the appropriate constant. Repeating the calculations recursively concludes the proof. $\square$

From the properties of the constructed sequence $(f_n)_{n \in \mathbb{N}}$, we can deduce an important property of the function $\hat{f}$.

PROPOSITION 5.2. *The function $\hat{f}$ belongs to the class $\mathcal{C}_e([0,T] \times (0,\infty)) \cap \mathcal{C}^{0,1}([0,T] \times (0,\infty))$. Moreover, its derivative satisfies*

$$(5.8) \qquad \left|\frac{\partial \hat{f}}{\partial y}(t,y)\right| \leq \phi(t) e^{A''(T-t)+B''y}, \qquad (t,y) \in [0,T] \times (0,\infty),$$

*for $A''$, $B''$ and $\phi$ as in Lemma 5.1.*

PROOF. The result follows if we show that the sequence $(\frac{\partial f_n}{\partial y})_{n \in \mathbb{N}}$, constructed in Lemma 5.1, converges uniformly, at least on compact subsets of $[0,T] \times (0,\infty)$.

Choose $n \geq m$ and $\rho > \alpha \vee A''$. Using the definition of the derivative (5.4), we have

$$e^{-\rho(T-t)-2B''y}\left|\frac{\partial f_{n+1}}{\partial y}(t,y) - \frac{\partial f_{m+1}}{\partial y}(t,y)\right|$$

$$\leq \mathbb{E}\bigg[(1-\gamma)\int_t^T \gamma e^{\gamma \int_t^s Q(Y^{t,y}(u))du}$$

$$\times |f_n(s,Y^{t,y}(s))^{-\gamma/(1-\gamma)}$$

$$- f_m(s,Y^{t,y}(s))^{-\gamma/(1-\gamma)}|$$

$$\times \int_t^s \left|\frac{\partial Q}{\partial y}(Y^{t,y}(u))\right| e^{-\lambda(u-t)}(u)\,du\,ds\bigg] e^{-\rho(T-t)-2B''y}$$

$$+ \mathbb{E}\bigg[\int_t^T \gamma e^{\gamma \int_t^s Q(Y^{t,y}(u))du} |f_n(s,Y^{t,y}(s))^{-1/(1-\gamma)}$$

(5.9)
$$- f_m(s,Y^{t,y}(s))^{-1/(1-\gamma)}|$$

$$\times \frac{\partial f_m}{\partial y}(s,Y^{t,y}(s))e^{-\lambda(s-t)}\,ds\bigg] e^{-\rho(T-t)-2B''y}$$

$$+ \mathbb{E}\bigg[\int_t^T \gamma e^{\gamma \int_t^s Q(Y^{t,y}(u))du}$$

$$\times \left|\frac{\partial f_n}{\partial y}(s,Y^{t,y}(s)) - \frac{\partial f_m}{\partial y}(s,Y^{t,y}(s))\right|$$

$$\times f_n(s,Y^{t,y}(s))^{-1/(1-\gamma)}e^{-\lambda(s-t)}\,ds\bigg] e^{-\rho(T-t)-2B''y}$$

$$=: M_1 + M_2 + M_3.$$

We first derive an upper bound for $M_1$. Again, let $d(\cdot,\cdot)$ denote the metric defined in (4.7). Applying the estimate (5.5) and the mean value theorem, we find

(5.10)
$$M_1 \leq \gamma^2 \mathbb{E}\bigg[\int_t^T e^{\gamma A(s-t)+B'(y+L(\lambda s)-L(\lambda t)-Y^{t,y}(s))}e^{\rho(T-s)+B'Y^{t,y}(s)}$$

$$\times e^{-\alpha(T-s)-B'Y^{t,y}(s)}$$

$$\times |f_n(s,Y^{t,y}(s)) - f_m(s,Y^{t,y}(s))|$$

$$\times \left(\frac{4D}{\gamma B'} \vee C\right) e^{(\gamma B'/4)(y+L(\lambda s)-L(\lambda t))} \int_t^s e^{-\lambda(u-t)}\,du\,ds\bigg]$$

$$\times e^{-\rho(T-t)-2B''y}$$

$$\leq \frac{1}{\lambda}\left(\frac{4D\gamma}{B'} \vee C\gamma^2\right) d(f_n,f_m) \int_t^T e^{(\gamma A+\lambda \psi(B'')-\rho)(s-t)}\,ds$$

$$\leq K_1 d(f_n,f_m).$$

Similarly, we have

$$M_2 \leq \frac{\gamma}{1-\gamma}\mathbb{E}\bigg[\int_t^T e^{\gamma A(s-t)+B''(y+L(\lambda s)-L(\lambda t)-Y^{t,y}(s))}e^{\rho(T-s)+B'Y^{t,y}(s)}$$

$$\times e^{-\alpha(T-s)-B'Y^{t,y}(s)}|f_n(s,Y^{t,y}(s)) - f_m(s,Y^{t,y}(s))|$$

$$\times \phi(s)e^{A''(T-s)+B''Y^{t,y}(s)}e^{-\lambda(s-t)}\,ds\bigg]$$

$$\times e^{-\rho(T-t)-2B''y}$$

$$(5.11) \quad \leq \frac{\gamma}{1-\gamma}\mathrm{d}(f_n,f_m)\mathbb{E}\bigg[\int_t^T e^{\gamma A(s-t)+B''(y+L(\lambda s)-L(\lambda t)-Y^{t,y}(s))}$$

$$\times e^{\rho(t-s)+B'(y+L(\lambda s)-L(\lambda t))}$$

$$\times \phi(s)e^{A''(T-s)+B''Y^{t,y}(s)}e^{-\lambda(s-t)}e^{-2B''y}\,ds\bigg]$$

$$\leq \frac{\gamma}{1-\gamma}\mathrm{d}(f_n,f_m)e^{A''T}\sup_{t\in[0,T]}\{\phi(t)\}\int_t^T e^{(\gamma A+\lambda\psi(2B'')-\rho-\lambda)(s-t)}\,ds$$

$$\leq K_2\mathrm{d}(f_n,f_m),$$

where we have used the bound (5.3) for the sequence of derivatives $(\frac{\partial f_n}{\partial y})_{n\in\mathbb{N}}$. Finally, we obtain a bound for $M_3$:

$$M_3 \leq \mathbb{E}\bigg[\int_t^T \gamma e^{\gamma A(s-t)+2B''(y+L(\lambda s)-L(\lambda t)-Y^{t,y}(s))}$$

$$(5.12) \quad \times e^{-\rho(T-s)-2B''Y^{t,y}(s)}\bigg|\frac{\partial f_n}{\partial y}(s,Y^{t,y}(s)) - \frac{\partial f_m}{\partial y}(s,Y^{t,y}(s))\bigg|$$

$$\times e^{\rho(T-s)+2B''Y^{t,y}(s)}e^{-\lambda(s-t)}\,ds\bigg]$$

$$\times e^{-\rho(T-t)-2B''y}$$

$$\leq \sup_{(t,y)\in[0,T]\times(0,\infty)}\bigg|e^{-\rho(T-t)-2B''y}\bigg(\frac{\partial f_n}{\partial y}(t,y) - \frac{\partial f_m}{\partial y}(t,y)\bigg)\bigg|$$

$$\times \gamma\int_t^T e^{(\gamma A+\lambda\psi(2B'')-\rho-\lambda)(s-t)}\,ds$$

$$\leq K_3\sup_{(t,y)\in[0,T]\times(0,\infty)}\bigg|e^{-\rho(T-t)-2B''y}\bigg(\frac{\partial f_n}{\partial y}(t,y) - \frac{\partial f_m}{\partial y}(t,y)\bigg)\bigg|,$$

where $\rho$ must be chosen such that $K_3 = \gamma(\rho - \gamma A - \lambda\psi(2B'') + \lambda)^{-1} < 1$.

Note that by the contraction property of the operator $\mathcal{L}$ proved in Lemma 4.2, we have

$$(5.13) \quad \begin{aligned}\mathrm{d}(f_n,f_m) &\leq \bigg(\frac{\gamma}{\alpha-A'}\bigg)^{m-1}\mathrm{d}(f_{n-m+1},f_1)\\ &\leq 2\bigg(\frac{\gamma}{\alpha-A'}\bigg)^{m-1}\bigg(1+\frac{1-\gamma}{A'}\bigg).\end{aligned}$$

By combining (5.10)–(5.13), we get

$$\left| e^{-\rho(T-t)-2B''y} \left( \frac{\partial f_{n+1}}{\partial y}(t,y) - \frac{\partial f_{m+1}}{\partial y}(t,y) \right) \right|$$

$$\leq 2(K_1+K_2)\left(\frac{\gamma}{\alpha-A'}\right)^{m-1}\left(1+\frac{1-\gamma}{A'}\right)$$

$$+ K_3 \sup_{(t,y)\in[0,T]\times(0,\infty)} \left| e^{-\rho(T-t)-2B''y} \left( \frac{\partial f_n}{\partial y}(t,y) - \frac{\partial f_m}{\partial y}(t,y) \right) \right|$$

$$\leq 2(K_1+K_2)\left(\frac{\gamma}{\alpha-A'}\right)^{m-1}\left(1+\frac{1-\gamma}{A'}\right)\frac{1-K_3^{m-1}}{1-K_3}$$

$$+ K_3^{m-1} \sup_{(t,y)\in[0,T]\times(0,\infty)} \left| e^{-\rho(T-t)-2B''y} \left( \frac{\partial f_{n-m+1}}{\partial y}(t,y) - \frac{\partial f_1}{\partial y}(t,y) \right) \right|$$

$$\leq 2(K_1+K_2)\left(\frac{\gamma}{\alpha-A'}\right)^{m-1}\left(1+\frac{1-\gamma}{A'}\right)\frac{1-K_3^{m-1}}{1-K_3} + 2K_3^{m-1} \sup_{t\in[0,T]}\{\phi(t)\},$$

from which we conclude that the sequence $(\frac{\partial f_n}{\partial y}(t,y))_{n\in\mathbb{N}}$ converges uniformly on compact sets. $\square$

We now turn to the question of differentiability in the time variable. We first show that the function $\hat{f}(t,y)$ belongs, for every fixed $t\in[0,T]$, to the domain of the infinitesimal generator of the process $Y$; see Chapter 1.3 in [19].

As the mapping $y \mapsto \hat{f}(t,y)$ is continuously differentiable on $(0,\infty)$, we can apply Itô's formula and show that the limit relation

$$\lim_{s\to 0} \frac{\mathbb{E}[\hat{f}(t,Y^{0,y}(s))] - \hat{f}(t,y)}{s}$$

(5.14)

$$= -\frac{\partial \hat{f}}{\partial y}(t,y)\lambda y + \int_{z>0} (\hat{f}(t,y+z) - \hat{f}(t,y))\nu(dz)$$

holds, provided that, for $s>0$,

$$(5.15) \quad \mathbb{E}\left[\int_0^s \int_{z>0} (\hat{f}(t,Y^{0,y}(u-)+z) - \hat{f}(t,Y^{0,y}(u-)))\tilde{N}(du\times dz)\right] = 0,$$

where $\tilde{N}(du\times dz) := N(du\times dz) - \nu(dz)\,du$ is the compensated Poisson random measure from (2.5). It is well known (see, e.g., Theorem 4.2.3 in [1]) that condition (5.15) is equivalent to

$$(5.16) \quad \mathbb{E}\left[\int_0^s \int_{z>0} |\hat{f}(t,Y^{0,y}(u-)+z) - \hat{f}(t,Y^{0,y}(u-))|^2 \nu(dz)\,du\right] < \infty.$$

The mean value theorem and the bound (5.8) imply that

$$\mathbb{E}\left[\int_0^s \int_{z>0} |\hat{f}(t, Y^{0,y}(u-) + z) - \hat{f}(t, Y^{0,y}(u-))|^2 \nu(dz)\, du\right]$$

$$\leq \mathbb{E}\left[\int_0^s \int_{z>0} \phi^2(t) e^{2A''(T-t)+2B''(Y^{0,y}(u)+z)} z^2 \nu(dz)\, du\right]$$

(5.17)

$$\leq \mathbb{E}\left[\int_0^s \int_{z>0} \phi^2(t) e^{2A''(T-t)+2B''y+2B''L(\lambda T)} e^{2B''z} z^2 \nu(dz)\, du\right]$$

$$\leq K e^{2B''} \int_{0<z<1} z^2 \nu(dz) + K \int_{z\geq 1} e^{2B''z} z^2 \nu(dz)$$

for some positive constant $K$, which is finite since $\psi(2B'') < \infty$. The first term in (5.17) is clearly finite. We show that the second term is also finite. By applying the inequality $z \leq \frac{4}{\gamma B'} e^{\frac{\gamma B'}{4}z}$ and assumption (B), we find that

$$\int_{z\geq 1} e^{2B''z} z^2 \nu(dz) \leq \left(\frac{4}{\gamma B'}\right)^2 \int_{z\geq 1} e^{(2B''+\gamma B'/2)z}\, dz$$

$$= \left(\frac{4}{\gamma B'}\right)^2 \int_{z\geq 1} e^{2B'(1+\gamma/2)z}\, dz < \infty.$$

Before stating the next lemma, we would like to remark that the mapping $(t,y) \mapsto \int_{z>0} (\hat{f}(t, y+z) - \hat{f}(t,y))\nu(dz)$ is continuous on $[0,T] \times (0,\infty)$. This follows from the inequality

$$|\hat{f}(t, y+z) - \hat{f}(t,y)| \leq \phi(t) e^{A''(T-t)+B''(y+z)} z$$

and Lebesgue's dominated convergence theorem.

PROPOSITION 5.3. *The function $\hat{f}$ satisfies the partial integro-differential equation*

(5.18)
$$0 = \frac{\partial \hat{f}}{\partial t}(t,y) - \frac{\partial \hat{f}}{\partial y}(t,y)\lambda y + \lambda \int_{z>0} (\hat{f}(t, y+z) - \hat{f}(t,y))\nu(dz)$$
$$+ \hat{f}(t,y)\gamma Q(y) + (1-\gamma)(\hat{f}(t,y))^{-\gamma/(1-\gamma)}, \qquad \hat{f}(T,y) = 1,$$

*in the classical sense. In particular, the mapping $(t,y) \mapsto \frac{\partial \hat{f}}{\partial t}(t,y)$ is continuous on $[0,T) \times (0,\infty)$.*

PROOF. The idea of the proof is similar to that of the proof of Proposition 5.5 in [5]. We will calculate the limit in (5.14) explicitly by using the representation of $\hat{f}$.

Consider a fixed $t \in [0, T)$. Note that by the time homogeneity of $Y$, the equivalent representation holds:

$$\hat{f}(t, y) := \mathbb{E}\bigg[e^{\gamma \int_s^{T-t+s} Q(Y(w)) dw}$$
$$+ (1-\gamma) \int_s^{T-t+s} e^{\gamma \int_s^u Q(Y(w)) dw}$$
$$\times (\hat{f}(u+t-s, Y(u)))^{-\gamma/(1-\gamma)} du \,|\, Y(s) = y\bigg],$$

for $s \geq 0$. Let $\sigma((Y^{0,y}(s))$ denote the $\sigma$-algebra generated by the random variable $Y^{0,y}(s)$ as defined in (2.7). We have that

$$\hat{f}(t, Y^{0,y}(s))$$
$$= \mathbb{E}\bigg[e^{\gamma \int_s^{T-t+s} Q(Y(w)) dw}$$
$$+ (1-\gamma) \int_s^{T-t+s} e^{\gamma \int_s^u Q(Y(w)) dw}$$
$$\times (\hat{f}(u+t-s, Y(u)))^{-\gamma/(1-\gamma)} \,|\, \sigma((Y^{0,y}(s))\bigg]$$

holds $\mathbb{P}$-a.s. Applying the law of iterated expectations, we obtain

$$\mathbb{E}[\hat{f}(t, Y^{0,y}(s))]$$
$$= \mathbb{E}\bigg[e^{\gamma \int_s^{T-t+s} Q(Y(w)) dw}$$
$$+ (1-\gamma) \int_s^{T-t+s} e^{\gamma \int_s^u Q(Y(w)) dw}$$
$$\times (\hat{f}(u+t-s, Y(u)))^{-\gamma/(1-\gamma)} du \,|\, Y(0) = y\bigg].$$

Now, consider the difference $\mathbb{E}[\hat{f}(t, Y^{0,y}(s))] - \hat{f}(t, y)$ for some $s > 0$ in the neighborhood of 0. By simple algebraic manipulations, we find

$$\frac{1}{s}(\mathbb{E}[\hat{f}(t, Y^{0,y}(s))] - \hat{f}(t, y))$$
$$= \mathbb{E}\bigg[(1-\gamma) \int_s^{T-t+s} (\hat{f}(u+t-s, Y^{0,y}(u)))^{-\gamma/(1-\gamma)} e^{\gamma \int_0^u Q(Y^{0,y}(w)) dw}$$
$$\times \frac{1}{s}(e^{-\gamma \int_0^s Q(Y^{0,y}(w)) dw} - 1) du\bigg]$$
$$- \frac{1}{s}\mathbb{E}\bigg[(1-\gamma) \int_0^s (\hat{f}(u+t-s, Y^{0,y}(u)))^{-\gamma/(1-\gamma)} e^{\gamma \int_0^u Q(Y^{0,y}(w)) dw} du\bigg]$$

$$+ \mathbb{E}\left[e^{\gamma \int_0^{T-t+s} Q(Y^{0,y}(w))dw} \frac{1}{s}(e^{-\gamma \int_0^s Q(Y^{0,y}(w))dw} - 1)\right]$$

$$+ \frac{1}{s}(\hat{f}(t-s,y) - \hat{f}(t,y))$$

$$=: M_1(s) + M_2(s) + M_3(s) + M_4(s).$$

Note that

$$e^{\gamma \int_0^s Q(Y^{0,y}(w))dw} \le e^{\gamma AT + B'y + B'L(\lambda T)},$$

$$\frac{1}{s}(1 - e^{-\gamma \int_0^s Q(Y^{0,y}(w))dw}) \le \sup_{s \ge 0}\{\gamma Q(Y^{0,y}(s))\}$$

$$\le \gamma AT + \gamma B \sup_{s \ge 0}\{Y^{0,y}(s)\}$$

$$\le \gamma AT + \gamma By + \gamma BL(\lambda T)$$

$$\le (\lambda \vee \gamma AT)e^{B'y + B'L(\lambda T)}$$

hold $\mathbb{P}$-a.s. for $0 < s \le T$. The above estimates ensure that we can apply Lebesgue's dominated convergence theorem to obtain the following limits:

$$\lim_{s \to 0} M_1(s) = -\gamma Q(y)\mathbb{E}\bigg[(1-\gamma)\int_0^{T-t} e^{\gamma \int_0^u Q(Y^{0,y}(w))du}$$
$$\times (\hat{f}(u+t,Y^{0,y}(u)))^{-\gamma/(1-\gamma)} du\bigg],$$

$$\lim_{s \to 0} M_2(s) = -(1-\gamma)(\hat{f}(t,y))^{-\gamma/(1-\gamma)},$$

$$\lim_{s \to 0} M_3(s) = -\gamma Q(y)\mathbb{E}[e^{\gamma \int_0^{T-t} Q(Y^{0,y}(w))dw}].$$

Moreover, $\lim_{s \to 0}(M_1(s) + M_3(s)) = -\gamma Q(y)\hat{f}(t,y)$ holds and by combining these calculations with (5.14), we arrive at

$$\lim_{s \to 0} M_4(s) = -\lim_{s \to 0} \frac{\hat{f}(t,y) - \hat{f}(t-s,y)}{s}$$

$$= -\frac{\partial \hat{f}}{\partial y}(t,y)\lambda y + \int_{z > 0}(\hat{f}(t,y+z) - \hat{f}(t,y))\nu(dz)$$

$$+ \hat{f}(t,y)\gamma Q(y) + (1-\gamma)(\hat{f}(t,y))^{-\gamma/(1-\gamma)}.$$

We conclude that the derivative $\frac{\partial \hat{f}}{\partial t}$ exists and that $\hat{f}$ satisfies the partial integro-differential equation (5.18). Moreover, the mapping $(t,y) \mapsto \frac{\partial \hat{f}}{\partial t}(t,y)$ is continuous on $[0,T) \times (0,\infty)$ by the continuity of all terms on the right-hand side of (5.18). $\square$

We can also conclude that the function $\hat{f}$ is the only classical solution of the partial integro-differential equation (3.14) as, for any such solution, the Feynman–Kac representation must hold; see [4] for a similar argument.

**6. Optimality of the solution.** We shall conclude with the following theorem, which states that our solution is indeed optimal.

THEOREM 6.1.  *Assume that the conditions* (A1)–(A3) *and* (B) *hold. Define the investment strategy*

$$(6.1) \quad \hat{\pi}(t) = \arg\max_{\pi \in [0,1]} \{\pi(\mu(Y(t-)) - r(Y(t-))) - \tfrac{1}{2}\pi^2(1-\gamma)\sigma^2(Y(t-))\}$$

*and the consumption rate*

$$(6.2) \qquad \hat{c}(t) = X^{\hat{c},\hat{\pi}}(t)(\hat{f}(t,Y(t-)))^{-1/(1-\gamma)},$$

*where the function $\hat{f}$ is the unique solution of the fixed point equation (4.8) in the space $\mathcal{C}^{1,1}([0,T) \times (0,\infty)) \cap \mathcal{C}_e([0,T] \times (0,\infty))$ given by*

$$f(t,y) = \mathbb{E}\bigg[e^{\gamma \int_t^T Q(Y^{t,y}(s))ds}$$
$$+ (1-\gamma)\int_t^T e^{\gamma \int_t^s Q(Y^{t,y}(u))du} f(s, Y^{t,y}(s))^{-\gamma/(1-\gamma)}\, ds\bigg]$$

*and $X^{\hat{c},\hat{\pi}}$ is the wealth process of the agent under $(\hat{c},\hat{\pi})$, defined as*

$$(6.3) \begin{aligned} X^{\hat{c},\hat{\pi}}(t) &= xe^{\int_0^t (\hat{\pi}(s)(\mu(Y(s-)) - r(Y(s-))) + r(Y(s-)) - (\hat{f}(s,Y(s-)))^{-1/(1-\gamma)})ds} \\ &\quad \times e^{-1/2\int_0^t (\hat{\pi}(s))^2 \sigma^2(Y(s-))ds + \int_0^t \hat{\pi}(s)\sigma(Y(s-))dW(s)}. \end{aligned}$$

*The pair $(\hat{c},\hat{\pi})$ is then the optimal strategy for the investment and consumption problem (3.2).*

The proof of the above theorem is based on a verification theorem stating the conditions which a candidate value function should satisfy in order to coincide with the optimal value function. Theorem 3.1 of [19] is an appropriate verification theorem for jump-diffusion processes. We can prove that our candidate solution satisfies all of its conditions. In particular, we would like to point out that assumption (B) is needed for proving uniform integrability of the value function. We want to emphasize that we have been able to prove the optimality of the strategy under the weaker integrability assumption (B) on the Lévy measure of $L$ than is required in [5].

Finally, we want to mention that in [11], a different verification theorem for jump-diffusion processes has been proven, one which requires substantially weaker conditions than those in Theorem 3.1 of [19]. However, this result requires the strategy to be càglàd (left continuous with right limits), in contrast to the weaker progressively measurable condition as in the present paper. Details can be obtained from the authors on request.

REMARK 6.2. Assume that we state our problem for a deterministic function $Y$. The candidate for the value function is again $\bar{V}(t,x) = x^\gamma \bar{f}(t)$ resulting in the ODE

$$\frac{d\bar{f}}{dt}(t) + \bar{f}(t)\gamma Q(Y(t)) + (1-\gamma)(\bar{f}(t))^{-\gamma/(1-\gamma)} = 0, \qquad \bar{f}(T) = 1,$$

which has the solution given by the fixed point equation

$$\bar{f}(t) = e^{\gamma \int_t^T Q(Y(s))ds} + (1-\gamma) \int_t^T e^{\gamma \int_t^s Q(Y(u))du} \bar{f}(s)^{-\gamma/(1-\gamma)} ds,$$

$$0 \leq t \leq T.$$

If $Y$ is a stochastic process, then the function $\bar{f}$, defined as above, involves a random path of the volatility process, so it depends on $\omega \in \Omega$. It is tempting to believe that the function $\mathbb{E}[\bar{f}(t,\omega) \mid Y(t) = y]$ solves the optimization problem (3.2).

However, taking the operator $\mathcal{L}$ as defined in (4.1), we calculate

$$\mathcal{L}\mathbb{E}^{t,y}[\bar{f}(t,\omega)]$$

$$= \mathbb{E}^{t,y}\left[e^{\gamma \int_t^T Q(Y(s,\omega))ds}\right.$$

$$\left. + (1-\gamma) \int_t^T e^{\gamma \int_t^s Q(Y(u,\omega))du} (\mathbb{E}[\bar{f}(s,\omega)|\mathcal{F}_s])^{-\gamma/(1-\gamma)} ds\right]$$

$$\leq \mathbb{E}^{t,y}\left[e^{\gamma \int_t^T Q(Y(s,\omega))ds}\right.$$

$$\left. + (1-\gamma) \int_t^T e^{\gamma \int_t^s Q(Y(u,\omega))du} \bar{f}(s,\omega)^{-\gamma/(1-\gamma)} ds\right]$$

$$= \mathbb{E}^{t,y}[\bar{f}(t,\omega)],$$

where the equality holds if and only if $(Y(t,\omega))_{0 \leq t \leq T}$ is independent of $\omega$, hence deterministic.

We conclude that the function $\mathbb{E}^{t,y}[\bar{f}(t,\omega)]$ does not satisfy the fixed point equation (4.8) and, as a result, it is not the solution to our optimization problem. The optimal value function and the optimal investment and consumption strategy are, different, as one might have expected.

EXAMPLE 6.3. We consider a financial model of Barndorff-Nielsen and Shephard type. More precisely, we choose the time horizon $T = 1$ and $\gamma =$ 0.75 for the exponent of the power utility function. Furthermore, in (2.1), we choose $\lambda = 1/6$, in (2.2), we take $r(y) = 0$ and in (2.3), we take $\mu(y) = 0.1 + 0.5y$ and $\sigma^2(y) = y$. Let the subordinator $L$ be a compound Poisson process with jumps of intensity 0.5 and exponentially distributed jump sizes with expectation $1/15$. We set the initial volatility level at $Y(0) = 0.2$, which equals the expected long-term volatility.

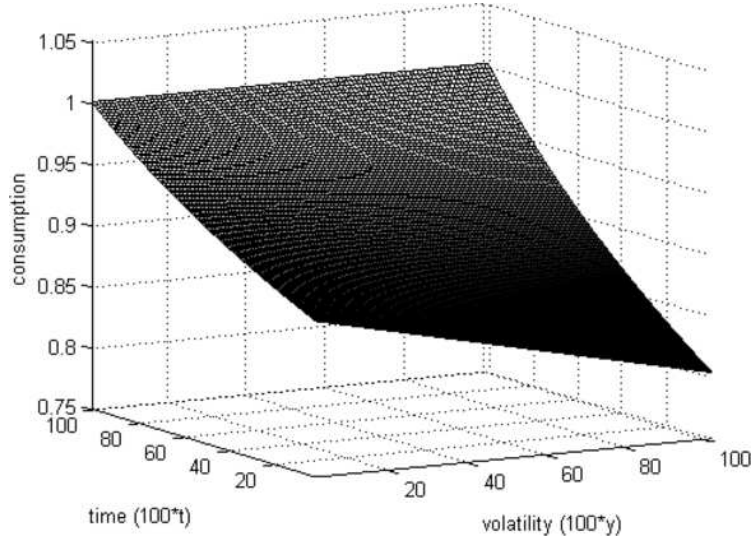

Fig. 1. *The optimal consumption rate as a function of time to maturity and the volatility level; see also the text in Example* 6.3.

We have solved the nonlinear partial integro-differential equation (3.14) numerically by applying an explicit finite difference method. As we are dealing with a first-order integro-differential equation and the Lévy measure is finite, the explicit scheme is more efficient than the implicit scheme; see [9], Chapter 12.4, for details. We point out that the finite difference method has been applied to the transformed equation to which the solution is $\hat{f}(t,y)e^{-\kappa y}$. The exponential scaling has been applied in order to set a sensible boundary condition in the bounded domain. The parameter $\kappa$ should be chosen sufficiently large so that $\lim_{y \to \infty} \hat{f}(t,y)e^{-\kappa y} = 0$ holds.

Based on (3.8), we can state that the optimal investment strategy is $\hat{\pi}(t) = 1$, whereas the optimal consumption rate is given in Figure 1.

The first observation, which is common in optimal investment and consumption models, is that the optimal consumption rate is an increasing function of time. In the model considered, it is interesting to note that the optimal consumption rate is a decreasing function of volatility level. The result agrees with our intuition: the higher level of volatility leads to a higher variability, which is, however, compensated for generously by an increase in the appreciation rate of the risky asset. This explains why the investor should consume less and invest more.

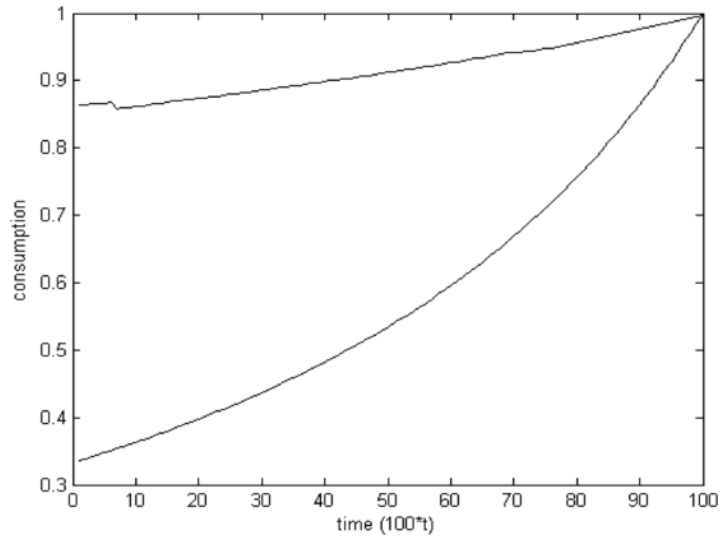

Fig. 2. *The optimal consumption rate in the stochastic volatility model* (*upper curve*) *and in the model with constant volatility* (*lower curve*); *see also the text in Example* 6.3.

It has already been stated, in [5], that stochastic volatility modeling can change the investment strategy significantly. We have simulated one path resulting in a volatility that jumps at $t = 0.05$ by $0.12$ and at $t = 0.65$ by $0.07$. The optimal consumption pattern is significantly different, when compared with the constant volatility model $Y \equiv 0.2$; see Figure 2. Under the stochastic volatility model, it is optimal to consume much higher proportions of the wealth as the unexpected jump in the volatility increases the variability of the return and may cause a severe decrease in the portfolio value. Note the discontinuity in the consumption strategy at $t = 0.05$ in the upper curve in Figure 2, which is caused by the jump in the volatility. The second discontinuity at $t = 0.65$ is much less visible.

We conclude this paper with the solution for the optimization problem in the case of a logarithmic utility. Logarithmic utility is investigated in depths in [14], where a general financial market is considered, consisting of stocks whose prices are driven by semimartingales. The solution is stated in terms of the semimartingales, characteristics, which are not straightforward to find in our model.

We make an ansatz with a value function of the form

$$v(t, x, y) = g(t) \log x + h(t, y).$$

This yields the following equations:

(6.4) $$0 = \frac{dg(t)}{dt} + 1, \qquad g(T) = 1$$

and

(6.5) $$\begin{aligned}0 = &\frac{\partial h}{\partial t}(t,y) - \frac{\partial h}{\partial y}(t,y)\lambda y \\ &+ \lambda \int_{z>0}(h(t,y+z) - h(t,y))\nu(dz) \\ &+ g(t)Q^0(y) - \log g(t) - 1 = 0, \qquad h(T,y) = 0,\end{aligned}$$

where

$$Q^0(y) = \max_{\pi \in [0,1]}\{\pi(\mu(y) - r(y)) - \tfrac{1}{2}\pi^2\sigma^2(y)\} + r(y)$$

is the analogue of (3.9).

The following theorem can be proven.

THEOREM 6.4. *Assume that conditions* (A1)–(A3) *hold and that the Lévy measure $\nu$ of $L$ satisfies the following condition:*

(C) $\int_{z>1} z^{1+\varepsilon}\nu(dz) < \infty$, *equivalently* $\mathbb{E}[L(1)^{1+\varepsilon}] < \infty$, *for some $\varepsilon > 0$.*

*Define the investment strategy*

$$\hat{\pi}(t) = \arg\max_{\pi \in [0,1]}\{\pi(\mu(Y(t-)) - r(Y(t-))) - \tfrac{1}{2}\pi^2\sigma^2(Y(t-))\}$$

*and the consumption rate*

$$\hat{c}(t) = \frac{X^{\hat{c},\hat{\pi}}(t)}{1 + T - t},$$

*where $X^{\hat{c},\hat{\pi}}$ is the wealth process of the agent under $(\hat{c},\hat{\pi})$, defined as*

$$X^{\hat{c},\hat{\pi}}(t) = xe^{\int_0^t (\hat{\pi}(s)(\mu(Y(s-)) - r(Y(s-))) + r(Y(s-)) - 1/(1+T-s))\,ds}$$
$$\times e^{-1/2\int_0^t (\hat{\pi}(s))^2 \sigma^2(Y(s-))ds + \int_0^t \hat{\pi}(s)\sigma(Y(s-))dW(s)}.$$

*The pair $(\hat{c},\hat{\pi})$ is then the optimal strategy for the investment and consumption problem under a logarithmic utility function.*

As we are facing the linear equation (6.5), existence and smoothness of a solution can be easily proven by combining the results from this paper with those and from [5]. It is well known (see [18]) that the optimal consumption rate in the case of a logarithmic utility does not depend on the financial coefficients.

**7. Conclusions.** In this paper, we have solved an investment and consumption problem for an agent who invests in a Black–Scholes market with stochastic coefficients driven by a non-Gaussian Ornstein–Uhlenbeck process. We have proven that the candidate value function is the classical solution of the corresponding Hamilton–Jacobi–Bellman equation. In particular, we have provided a classical solution to a nonlinear first-order partial integro-differential equation.

The optimal investment strategy has been explicitly calculated, while the optimal consumption rate depends on the function which solves the partial integro-differential equation. The conclusion from the simulation study is that under stochastic volatility, the optimal consumption strategy is significantly different compared to a constant volatility model.

In [5], a multivariate Ornstein–Uhlenbeck process driven by independent subordinators was considered, while in [17], a financial market consisting of $n$ stocks was investigated. We would like to point out that our results can be extended to both settings.

**Acknowledgments.** This paper was written while the first author was staying at Munich University of Technology. Łukasz Delong would like to thank Claudia Klüppelberg and the colleagues from the Department of Mathematics in Munich for their hospitality during his visit.

INSTITUTE OF ECONOMETRICS  
DIVISION OF PROBABILISTIC METHODS  
WARSAW SCHOOL OF ECONOMICS  
AL. NIEPODLEGLOSCI 162  
02-554 WARSAW  
POLAND  
E-MAIL: lukasz.delong@sgh.waw.pl

CENTER FOR MATHEMATICAL SCIENCES  
MUNICH UNIVERSITY OF TECHNOLOGY  
BOLTZMANNSTRASSE 3  
D-85747 GARCHING  
GERMANY  
E-MAIL: cklu@ma.tum.de